\documentclass{aa}
\usepackage{graphicx}
\usepackage{natbib}
\usepackage{amssymb}
\usepackage{amsfonts}
\usepackage{amsbsy}
\usepackage{amsmath}

\bibpunct{(}{)}{;}{a}{}{,}

\newcommand{\Scmbi}{\mathfrak{S}^{(i)}_{\rm c}(p)}
\newcommand{\Scmb}{\mathfrak{S}_{\rm c}(p)}
\newcommand{\Sbb}{\boldsymbol{\mathfrak{S}}}
\newcommand{\Scbj}{\boldsymbol{\mathfrak{S}}_j}

\newcommand{\Scba}{\boldsymbol{\mathfrak{S}}_1}
\newcommand{\Scbb}{\boldsymbol{\mathfrak{S}}_2}
\newcommand{\ScbNc}{\boldsymbol{\mathfrak{S}}_{Nc}}
\newcommand{\Scmbh}{\widehat{\mathfrak{S}}_{\rm c}(p)}
\newcommand{\Scmbhb}{\widehat{\boldsymbol{\mathfrak{S}}}_{\rm c}}
\newcommand{\Scmbb}{\boldsymbol{\mathfrak{S}}_{\rm c}}

\newcommand{\Smthb}{\boldsymbol{\mathfrak S}}

\newcommand{\Sgalb}{\boldsymbol{S}_{\rm f}}
\newcommand{\Sgalib}{\boldsymbol{S}^{(i)}_{\rm f}}
\newcommand{\Sib}{\boldsymbol{S}^{(i)}}
\newcommand{\Sgal}{S^{(i)}_{\rm f}(p)}
\newcommand{\Si}{S^{(i)}(p)}

\newcommand{\Cmthb}{\boldsymbol{\mathcal C}}

\newcommand{\Nmthi}{{\mathcal N}^{(i)}}
\newcommand{\Nmthb}{\boldsymbol{\mathcal N}}

\newcommand{\oneb}{\boldsymbol{1}}
\newcommand{\Ab}{\boldsymbol{A}}
\newcommand{\Fb}{\boldsymbol{F}}
\newcommand{\eb}{\boldsymbol{e}}
\newcommand{\Cb}{\boldsymbol{C}}
\newcommand{\Db}{\boldsymbol{D}}
\newcommand{\Hb}{\boldsymbol{H}}
\newcommand{\Ib}{\boldsymbol{I}}

\newcommand{\deltab}{\boldsymbol{\delta}}
\newcommand{\gammab}{\boldsymbol{\gamma}}
\newcommand{\Deltab}{\boldsymbol{\Delta}}
\newcommand{\Omegab}{\boldsymbol{\Omega}}

\newcommand{\Thetab}{\boldsymbol{\Theta}}

\newcommand{\Qb}{\boldsymbol{Q}}
\newcommand{\Sb}{\boldsymbol{S}}

\newcommand{\Vb}{\boldsymbol{V}}
\newcommand{\Zb}{\boldsymbol{Z}}
\newcommand{\wb}{\boldsymbol{w}}

\begin{document}

   \title{A Statistical Analysis of the \\ ``Internal Linear Combination'' Method \\ in Problems of Signal Separation \\ as in
   CMB Observations}

   \author{R. Vio\inst{1}
    \and
           P. Andreani\inst{2}
          }
   \institute{Chip Computers Consulting s.r.l., Viale Don L.~Sturzo 82,
              S.Liberale di Marcon, 30020 Venice, Italy\\
              \email{robertovio@tin.it},
         \and
		  ESO, Karl Schwarzschild strasse 2, 85748 Garching, Germany\\
                  INAF-Osservatorio Astronomico di Trieste, via Tiepolo 11, 34143 Trieste, Italy\\              
		  \email{pandrean@eso.org}
             }

\date{Received .............; accepted ................}

\abstract
{}
{The separation of foreground contamination from cosmic microwave background (CMB) observations is one of the most challenging and important problem
of digital signal processing in Cosmology. In literature, various techniques have been presented, but no general consensus about 
their real performances and properties has been reached. This is due to the characteristics of these  techniques that have been 
studied essentially through numerical simulations based on semi-empirical models of the CMB and the Galactic foregrounds. Such models often have 
different level of sophistication and/or are based on different physical assumptions (e.g., the number of the Galactic components and the
level of the noise). Hence, a reliable comparison is difficult.
What actually is missing is a statistical analysis of the properties of the proposed methodologies.
Here, we consider the {{\it Internal Linear Combination}} method (ILC) which, among the separation techniques,
requires the smallest number of {{\it a priori}} assumptions. This feature is of particular interest in the 
context of the CMB polarization measurements at small angular scales where
the lack of knowledge of the polarized backgrounds represents a serious limit.}
{The statistical characteristics of ILC are examined through an analytical approach and the basic conditions are fixed in a way to
work satisfactorily. A comparison with the FastICA implementation of the {{\it Independent Component Analysis}} (ICA) method, one
of the most celebrated techniques for blind signal separation, is made.}
{ILC provides satisfactory results only under rather restrictive conditions. This is a critical fact to take into consideration in planning
the future ground-based observations (e.g., with ALMA) where, contrary to the satellite experiments, there is the possibility to have a certain 
control of the experimental conditions.}
{}
\titlerunning{ILC method}
\authorrunning{R. Vio, \& P. Andreani}
\maketitle

\section{INTRODUCTION}

The experimental progresses in the detection of cosmological and astrophysical emissions require a 
parallel development of data analysis techniques in order to extract the maximum physical information 
from data. An example of very interest is represented by signals that are the mixture of the emission
of distinct physical mechanisms. The study of the underlying physical processes needs the separation of the
different components that contribute to the observed signals. This is the case of the Cosmic Microwave Background (CMB) observations where 
there is the necessity
to separate the CMB from diffuse foregrounds originated by our own Galaxy. In this context, an extensive literature is available
and many approaches have been proposed \citep[for a review, see][]{del07}. However, no general consensus about 
their real performances and properties has been reached.  This is due to the approach followed to determine the characteristics of these  techniques 
that is based on numerical simulations that make use of semi-empirical models of the CMB and the Galactic foregrounds. The point is that such models 
often have different level of sophistication and/or are based on different physical assumptions (e.g., the number of the Galactic components and the
level of the noise). Hence, a reliable comparison is difficult. 
A trustworthy assessment of the real capabilities of such methodologies should require a rigorous analysis of their theoretical 
statistical characteristics independently of the specific context where they are applied.
However, at least in our knowledge, in literature nothing has never been presented in this sense \citep[however, see][~for a discussion concerning
the power spectrum]{sah07}. 
Things become even more serious in the case of CMB polarization 
measurements, where the available {\it a priori} information is quite limited and the use of {\it blind} separation techniques obligatory \citep{sti06}.

A situation as this one is clearly unsatisfactory. This also because in a near future some innovative ground-based experiments are planned for polarization 
observations at extraordinary high spatial resolution as with the Atacama Large submmillimetre/millimetre array (ALMA). One important
advantage of these experiments is that, contrary to the satellite observations, they will allow a certain control of the experimental conditions. 
Hence, a fully exploitation of the capabilities of the instruments implies a careful preparation of the observations in such a way the obtained data 
are of sufficient quality to permit an effective application of the chosen separation methodology. For this reason, in this work we start exploring 
the capabilities of algorithms aimed at a careful subtraction of the foreground sources when the
amount of available {\it a priori} information is limited (an expected situation for polarization observations). In particular, 
we consider one of the most used approaches to the separation of different emissions, the {\it internal linear combination} method (ILC), 
which was adopted for instance in the reduction of the data from the Wilkinson Microwave Anisotropy Probe (WMAP) satellite for CMB observations 
\citep{ben03}. Among the separation techniques, ILC requires the smallest number of {\it a priori} assumptions.
For the reason presented above, we do not perform an application to any astrophysical dataset here, but rather we study the general properties of 
this technique in order to fix the conditions under which it can be expected to produce reliable results.
We also make a critical comparison with the architecture and capabilities of the FastIca implementation of the ICA approach, which is the other 
main handle to the separation when little {\it a priori} information is available \citep{sti06}. 

In the following, the available data are assumed in form of $N_o$ maps, taken at different frequencies, containing $N_p$ pixels each.  
More precisely, if $\Si$ provides the value of the $p$th pixel for a map obtained at channel ``$~i~$'' \footnote{In the present work, $p$ 
indexes pixels in the classic spatial domain. However, the same formalism applies if other domains are considered as, for example, the Fourier one.}, 
our starting model is:
\begin{equation} \label{eq:image}
\Si = \Scmbi + \Sgal + \Nmthi(p)
\end{equation}
where $\Scmbi$, $\Sgal$ and $\Nmthi(p)$ are the contributions due to the CMB, the diffuse Galactic foreground and the experimental noise, 
respectively.  Although not necessary for later arguments, it is assumed that all of these contributions are representable by means of stationary 
random fields. At least locally, in many experimental situations this is an acceptable approximation. If not, in any case
it is often made since it permits a statistical treatment of the problem of interest and the results can
be used as benchmark in the analysis of more complex scenarios. In the present context, this assumption holds on small
patches of the sky. Finally, without loss of generality, for easiness of notations the random fields are supposed the
realization of zero-mean spatial processes. In the present work the contribution of non-diffuse components 
(e.g., due to SZ cluster, point-sources, \ldots) is not considered and it is supposed to be already removed through other methodologies.

The paper is organized as follows: in Sec.~\ref{sec:formalization} the relevant analytical formulas are introduced. In Sec.~\ref{sec:ILC}
the statistical framework of ILC in noiseless observations is presented and compared with that derived for FastICA. Simulations are also
presented. In Sec.~\ref{sec:ILCnoise} the
case of noisy observations is considered. Finally, the conclusions are drawn in Sec.~\ref{sec:conclusions}.

\section{FORMALIZATION OF THE PROBLEM}
\label{sec:formalization}

The idea behind ILC is rather simple. The main assumption is that model~(\ref{eq:image}) can be written as
\begin{equation} \label{eq:basicm}
\Si = \Scmb + \Sgal + \Nmthi(p),
\end{equation}
i.e. the template of the CMB component is independent of the observing channel. A natural idea to exploit this fact is to 
average $N_o$ images $\{ \Si \}_{i=1}^{N_o}$ giving a specific weight $w_i$ to each of them in such a way to minimize the 
impact of the foregrounds and noise
\citep{ben03}. This means to look for a solution of type
\begin{equation} \label{eq:sum}
\Scmbh = \sum_{i=1}^{N_o} w_i \Si.
\end{equation}
In fact, if the constraint $\sum_{i=1}^{N_o} w_i = 1$ is imposed, Eq.~(\ref{eq:sum}) becomes
\begin{equation} \label{eq:wls}
\Scmbh = \Scmb + \sum_{i=1}^{N_o}  w_i [ \Sgal + \Nmthi(p) ]. 
\end{equation}
Now, from this equation it is clear that, for a given pixel ``$p$'',  the only variable terms are in the summatory. Hence,  
under the assumption of independence of $\Scmb$ from $\Sgal$ and $\Nmthi(p)$,
the weights $\{ w_i \}$ have to minimize the variance of $\Scmbh$, i.e.
\begin{align}
& \{ w_i \} = \underset{\{ w_i \} }{\arg\min} \nonumber \\
& {\rm VAR} \left[\Scmb \right] + {\rm VAR}\left[\sum_{i=1}^{N_o}  w_i (\Sgal + \Nmthi(p)) \right],
\end{align}
where ${\rm VAR}[s(p)]$ is the {\it expected variance} of $s(p)$.
If $\Sib$ denotes a {\bf row} vector such as $\Sib = [S^{(i)}(1), S^{(i)}(2), \ldots, S^{(i)}(N_p)]$ and the $N_o \times N_p$ matrix $\Sb$ is defined as  
\begin{equation}
\Sb = 
\left( \begin{array}{c}
\Sb^{(1)} \\
\Sb^{(2)} \\
\vdots \\
\Sb^{(N_o)}
\end{array} \right),
\end{equation}
then Eq.~(\ref{eq:basicm}) becomes
\begin{equation}
\Sb = \Scmbb + \Sgalb + \Nmthb.
\end{equation}
In this case, the weights are given by \citep{eri04}
\begin{equation} \label{eq:wr}
\wb = \frac{\Cb_{\Sb}^{-1} \oneb}{\oneb^T \Cb_{\Sb}^{-1} \oneb},
\end{equation}
where 
$\Cb_{\Sb}$ is the $N_o \times N_o$ cross-covariance matrix of the random processes that generate $\Sb$, i.e.
\begin{equation}
\Cb_{\Sb} = {\rm E}[\Sb \Sb^T], 
\end{equation}
and $\oneb = (1, 1, \ldots, 1)^T$ is a column vector of all ones. Here, ${\rm E}[.]$ denotes the {\it expectation operator}.  
Hence, the ILC estimator takes the form
\begin{align} 
\Scmbhb & = \wb^T \Sb, \label{eq:wlss} \\
        & = \alpha \oneb^T \Cb_{\Sb}^{-1} \Sb, \label{eq:basic}
\end{align}
with 
$\oneb^T \wb = 1$ and the scalar quantity $\alpha$ given by
\begin{equation} \label{eq:alpha}
\alpha = [ \oneb^T \Cb_{\Sb}^{-1} \oneb]^{-1}.
\end{equation}

In practical applications, matrix $\Cb_{\Sb}$ is unknown and has to be estimated from the data. Typically, this is done by means of the
estimator
\begin{equation} \label{eq:C}
\widehat \Cb_{\Sb} = \frac{1}{N_p} \Sb \Sb^T.
\end{equation}
In this case, the ILC estimator is given by Eqs.(\ref{eq:wlss})-(\ref{eq:alpha}) with $\Cb_{\Sb}$ and $\wb$ replaced, respectively, by 
$\widehat \Cb_{\Sb}$ and
\begin{equation} \label{eq:w}
\widehat \wb = \frac{\widehat \Cb_{\Sb}^{-1} \oneb}{\oneb^T \widehat \Cb_{\Sb}^{-1} \oneb}.
\end{equation}
Here, it is important to underline that if the observed maps are not zero-mean, they have to be centered before the computation of $\widehat \Cb_{\Sb}$. 
After that, the resulting weights can be applied directly to the original (i.e., non-centered) $\Sb$. The computation of $\widehat \Cb_{\Sb}$ is
the only point where the fact of working with non-zero mean maps has to be taken into account.

Although in the literature it might appear that the estimator~(\ref{eq:basic}) has {\it optimal} 
properties, actually this is not true. The point is that in
Eq.~(\ref{eq:basicm}) the term $\Sgal + \Nmthi(p) $ is considered as a single noise component 
\citep[e.g., see ][]{eri04,hin07}. 
In this way the problem is apparently simplified since it is reduced to the separation of two components only. No {\it a priori} information
on this ``global'' noise is required. However, this approach can lead to wrong 
conclusions. For example, since all the components in the mixtures $\Sb$ are assumed 
to be zero-mean, from Eq.~(\ref{eq:wls}) one could conclude that
\begin{equation}
{\rm E}[\Scmbhb | \Scmbb] = \Scmbb + \wb^T{\rm E}[\Sb] = \Scmbb,
\end{equation}
i.e. the ILC estimator is unbiased \footnote{The expression ${\rm E}[a|b]$ indicates {\it conditional expectation}
of $a$ given $b$. }. This is not correct: the claim that $\Scmbhb$ is unbiased requires to prove that
\begin{equation}
{\rm E}[\Scmbhb | \Scmbb, \Sgalb] = \Scmbb + \wb^T \Sgalb + \wb^T {\rm E}[\Nmthb] = \Scmbb.
\end{equation}
The reason is that $\Sgalb$ is a fixed realization of a random process. There is no possibility to get another one. 
Even if observed many times (under the same experimental conditions) the Galactic components will appear always the same.
Only the noise component $\Nmthb$ will change. This has important consequences. In order to discuss this issue, it is useful to start 
with the case of noiseless observations that can be thought to reproduce a situation of very high signal-to-noise ratio. In this case, 
model~(\ref{eq:basicm}) becomes
\begin{equation} \label{eq:imagenonoise}
\Sib = \Scmbb + \Sgalib.
\end{equation}

\section{STATISTICAL CHARACTERISTICS OF ILC: NOISELESS OBSERVATIONS}
\label{sec:ILC}

A common assumption in CMB observations is that $\Sgalib$ is given by the linear mixture of the contribution of 
$N_c$ physical processes $\{ \Scbj \}_{j=1}^{N_c}$ (e.g., free-free, dust re-radiation, 
\ldots)
\begin{equation} \label{eq:gal}
\Sgalib = \sum_{j=1}^{N_c} a_{ij} \Scbj,
\end{equation}
with $a_{ij}$ constant coefficients. In practice, it is assumed that for the $j$th physical process a template $\Scbj$ exists
independent of the specific channel ``$~i~$''. Although rather strong, actually it is not unrealistic to assume that this condition is satisfied 
when small enough patches of the sky are considered. Inserting Eq.~(\ref{eq:gal}) into Eq.~(\ref{eq:imagenonoise}) one obtains
\begin{equation} \label{eq:model}
\Sb = \Ab \Smthb,
\end{equation}
with
\begin{equation}
\Smthb = \left( \begin{array}{l}
\Scmbb \\
\Scba \\
\Scbb \\
\vdots \\
\ScbNc \\
\end{array} \right),
\end{equation}
and
\begin{equation}
\Ab = 
\left( \begin{array}{cccccc} \label{eq:Amatrix}
1 & \vline & a_{11} & a_{12} & \ldots & a_{1 N_c} \\
\hline
1 & \vline & a_{21} & a_{22} & \ldots & a_{2 N_c} \\
\vdots & \vline & \vdots & \vdots & \ddots & \vdots \\
1 & \vline & a_{N_o 1} & a_{N_o 2} & \ldots & a_{N_o N_c}
\end{array} \right).
\end{equation}
Here, matrix $\Ab$, assumed to be of full rank, is shown partitioned in a way that will be useful for later calculations.
 
In the following, three cases are considered that correspond to possible experimental situations. 

\subsection{Case $N_o = N_c + 1$} \label{sec:equal}

If $N_o = N_c + 1$, i.e. when number of observations is equal to the number of the components (CMB included),
then $\Ab$ is a square $N_o \times N_o$ matrix. In this case,
\begin{equation} \label{eq:cov}
\Cb_{\Sb} = \Ab \Cb_{\Smthb} \Ab^T
\end{equation}
with $\Cb_{\Smthb}$ the $N_o \times N_o$ cross-covariance matrix of the random processes that generate the templates $\Smthb$, i.e.
\begin{equation}
\Cb_{\Smthb} = {\rm E}[\Smthb \Smthb^T].
\end{equation}
If this equation is inserted in Eq.~(\ref{eq:basic}), one obtains
\begin{equation} \label{eq:solution}
\Scmbhb = \alpha \oneb^T \Ab^{-T} \Cb_{\Smthb}^{-1} \Smthb,
\end{equation}
with the scalar $\alpha$ given by
\begin{equation}
\alpha = [ \oneb^T \Ab^{-T} \Cb_{\Smthb}^{-1} \Ab^{-1} \oneb]^{-1},
\end{equation}
and $\Ab^{-T} \equiv (\Ab^{-1})^T$.
Now, since it is trivially verified that 
\begin{equation}
\oneb^T = \eb_1^T \Ab^T
\end{equation}
with
\begin{equation} \label{eq:trick2}
\eb_1 \equiv (1, 0, \ldots, 0)^T,
\end{equation}
it is
\begin{equation} \label{eq:trick1}
\oneb^T \Ab^{-T} = \eb_1^T.
\end{equation}
Hence, $\alpha = (\Cb_{\Smthb}^{-1})_{1 1} = ({\rm E}[\Scmbb \Scmbb^T])^{-1} = \sigma_{\rm cc}^{-1}$ is the inverse of the expected variance of the 
CMB template. 
As a consequence, if the random process that generates the template $\Scmbb$ is uncorrelated with those that generate the templates $\{ \Scbj \}$ 
(in general, this is a reasonable assumption), i.e. if
\begin{equation}
\Cb_{\Smthb} = 
\left( \begin{array}{cccccc} \label{eq:covar1}
\sigma_{\rm cc} & \vline & 0 & 0 & \ldots & 0 \\
\hline
0 & \vline & \sigma_{11} & \sigma_{12} & \ldots & \sigma_{1 N_o} \\
\vdots & \vline & \vdots & \vdots & \ddots & \vdots \\
0 & \vline & \sigma_{N_o 1} & \sigma_{N_o 2} & \ldots & \sigma_{N_o N_o}
\end{array} \right),
\end{equation}
from Eq.~(\ref{eq:solution}) and the fact that $\Cb_{\Smthb}^{-1}$ has the form
\begin{equation}
\Cb^{-1}_{\Smthb} = 
\left( \begin{array}{cccccc}
\sigma^{-1}_{\rm cc} & \vline & 0 & 0 & \ldots & 0 \\
\hline
0 & \vline & & & & \\
\vdots & \vline & & & \Omegab^{-1} & \\
0 & \vline & & & &
\end{array} \right),
\end{equation}
with $\Omegab$ the bottom-right block of the matrix in the rhs of Eq.~(\ref{eq:covar1}), one obtains that 
\begin{equation}
\Scmbhb = \Scmbb.
\end{equation}
Here, no use is made of the operator ${\rm E}[.|.]$ since we are dealing with fixed realizations of the random processes that
generate the CMB as well the Galactic components. Therefore, if matrix $\Cb_{\Sb}$ is known, the ILC solution is exact. 

This condition changes if matrix $\Cb_{\Sb}$ has to be estimated through Eq.~(\ref{eq:C}) and a general treatment of this problem is quite difficult. 
For this reason, we consider the case of observations that span a sky area much wider than the correlation lengths of the maps $\{ \Sib \}$. Under this
condition, $\widehat \Cb_{\Smthb} = \Smthb \Smthb^T/ N_p$ can be written in the form
\begin{equation}
\widehat \Cb_{\Smthb} = \Cb_{\Smthb} + \Deltab \Cb_{\Smthb}
\end{equation}
with
\begin{equation}
\Deltab C_{\Smthb} =
\left( \begin{array}{cccccc} \label{eq:delcovar}
\delta_{\rm cc} & \vline & \delta_{c1} & \delta_{c2} & \ldots & \delta_{c N_o}  \\
\hline
\delta_{c1} & \vline & \delta_{11} & \delta_{12} & \ldots & \delta_{1 N_o}  \\
\vdots & \vline & \vdots & \vdots & \ddots & \vdots \\
\delta_{c N_o} & \vline & \delta_{1 N_o} & \delta_{2 N_o} & \ldots & \delta_{N_o N_o}
\end{array} \right)
\end{equation}
a perturbing matrix with small zero-mean entries. Because of this, it can be expanded in Taylor 
series around $\Cb_{\Smthb}^{-1}$ 
up to the linear term obtaining
\begin{equation} \label{eq:taylor}
\widehat \Cb_{\Smthb}^{-1} \approx  \Cb_{\Smthb}^{-1} - \Cb_{\Smthb}^{-1} \Deltab \Cb_{\Smthb} \Cb_{\Smthb}^{-1},
\end{equation}
or
\begin{equation}
\widehat \Cb_{\Smthb}^{-1}
\approx \left( \begin{array}{ccc}
\sigma^{-1}_{\rm cc} - \sigma^{-2}_{\rm cc} \delta_{\rm cc} & \vline & - \sigma^{-1}_{\rm cc} \deltab^T \Omegab^{-1} \\
\hline
-\sigma^{-1}_{\rm cc} \Omegab^{-1} \deltab & \vline & \Omegab^{-1} - \Omegab^{-1} \Thetab \Omegab^{-1}
\end{array} \right),
\end{equation}
with $\deltab = (\delta_{c1}, \delta_{c2}, \ldots, \delta_{c N_o})^T$ and $\Thetab$ the bottom-right block of the matrix in the rhs 
of Eq.~(\ref{eq:delcovar}) \footnote{It is necessary to stress that the entries of
$\Deltab \Cb_{\Smthb}$ are not independent. The point is that if $\Cb$ is a symmetric, positive definite matrix, then the same holds for 
its inverse $\Cb^{-1}$. However, in general, this is not true for a matrix $\Cmthb = \Cb + \Deltab \Cb$ obtained perturbing $\Cb$ with an arbitrary
matrix. This means that the entries of 
$\Deltab \Cb$ have to satisfy certain conditions. If $\Cmthb $ is written in the form
$ [(\Qb + \Deltab \Qb) (\Qb + \Deltab \Qb)^T]$, with $\Qb = \Cb^{1/2}$ and $\Deltab Q$ a matrix of zero-mean random quantities, and then expanded 
expanded in Taylor series, one can find the same result as in 
Eq.~(\ref{eq:taylor}) with $\Deltab \Cb = (\Deltab \Qb)  \Qb^T + \Qb (\Deltab \Qb)^T $.}.
From this result and from Eqs.~(\ref{eq:solution}), (\ref{eq:trick2}) and (\ref{eq:trick1}), one obtains that
\begin{equation} \label{eq:delta1}
\Scmbhb = (1, \deltab_{\Sbb}) \Smthb,
\end{equation}
or
\begin{equation}
\Delta \Scmbb = \Scmbhb - \Scmbb = (0, \deltab_{\Sbb}) \Smthb,
\end{equation}
with $\deltab_{\Sbb}$ a row vector given by
\begin{equation} \label{eq:delta2}
\deltab_{\Sbb} = - \frac{\sigma_{\rm cc}}{\sigma_{\rm cc} - \delta_{\rm cc}} \deltab^T \Omegab^{-1} \approx
- \deltab^T \Omegab^{-1}.
\end{equation}
Hence, not unexpectedly, the ILC solution differs from the true one by an amount that depends on the sample correlation between 
$\Scmbb$ and the Galactic templates.

\subsection{Case $N_o < N_c + 1$} \label{sec:minus}

If $N_o < N_c + 1$ , the number of the components (CMB included) is larger than the number of channels. Since in this case $\Ab$ is a rectangular
$N_o \times (N_c+1)$ matrix, the inverse $\Ab^{-1}$ is not defined. Hence, the ILC solution $\Scmbhb$ as given by Eq.~(\ref{eq:basic}) cannot be 
written in the form~(\ref{eq:solution}). The only possibility for still having $\Scmbhb = \Scmbb$ is that 
\begin{equation}
\alpha \oneb^T  ( \Ab \Cb_{\Smthb}^{-1} \Ab^T )^{-1} \Ab = \eb_1^T.
\end{equation}
In the present case, there is no reason to expect that this condition be satisfied. As a consequence, the ILC solution differs from the true one
by an amount given by 
\begin{equation} \label{eq:bias}
\Delta \Scmbb = [\alpha \oneb^T (\Ab \Cb_{\Smthb} \Ab^T)^{-1} \Ab - \eb_1^T] \Smthb.
\end{equation}
This result implies that, similarly to other techniques (e.g., Generalized Least Squares), it is risky to use the ILC technique in situations where
the number of observing channels is smaller than the number of components. Unfortunately, the importance of $\Delta \Scmbb$ depends on the specific 
characteristics of both $\Ab$ and $\Cb_{\Smthb}$ and, hence, a general treatment is not possible. However, it is not difficult to realize that 
when the Galactic component is the dominant one, it can even happen that $\Delta \Scmbb > \Scmbb$. For example, 
in the hypothesis that $\Cb_{\Smthb} = \sigma^2 \Ib$ and
\begin{equation}
\Ab = 
\left( \begin{array}{ccc}
1.0 & 2.0 & 3.0 \\
1.0 & 2.5 & 3.5
\end{array} \right),
\end{equation}
Eq.~(\ref{eq:bias}) provides
\begin{equation}
\Delta \Scmbb = (-0.33, -0.33, 0.33) \Smthb.
\end{equation}
This example does not illustrate a situation excessively unfavorable for the separation. In fact, close to the Galactic 
plane the CMB component is expected to be largely dominated by the Galactic emissions.

\subsection{Case $N_o > N_c + 1$} \label{sec:more}

If $N_o > N_c + 1$, the number of components (CMB included) is smaller than the number of channels. As in the previous section, also
in this case $\Ab$ is $N_o \times (N_c + 1)$ rectangular matrix but now we face a more difficult situation. In fact, the ILC solution $\Scmbhb$ 
cannot be written in the form~(\ref{eq:basic}) since matrix $\Cb_{\Sb}$ is singular. The only way out is to resort to the {\it pseudo-inverse} 
$\Cb_{\Sb}^{\dag}$
\begin{equation}
\Cb_{\Sb}^{\dag} = \Vb \Db^{\dag} \Vb^T.
\end{equation}
Here, $\Vb$ is the orthogonal matrix obtained from the {\it singular value decomposition} of $\Cb_{\Sb}$,
\begin{equation} \label{eq:svd}
\Cb_{\Sb} = \Vb \Db \Vb^T,
\end{equation}
$\Db^{\dag}$ is a diagonal matrix whose entries are given by $d^{\dag}_{ii} = d_{ii}^{-1}$ if $d_{ii} \neq 0$, 
$0$ otherwise, and 
$\{d_{ii} \}$ are the elements of the diagonal matrix $\Db$. In this case, the ILC solution is given by
\begin{equation} \label{eq:basic1}
\Scmbhb = \alpha \oneb^T \Cb_{\Sb}^{\dag} \Sb, 
\end{equation}
where
\begin{equation} \label{eq:alpha1}
\alpha = [ \oneb^T \Cb_{\Sb}^{\dag} \oneb]^{-1}.
\end{equation}
If $\Ab$ is a full column-rank matrix, then Eq.~(\ref{eq:basic1}) can be rewritten in the form
\begin{equation} \label{eq:solution1}
\Scmbhb = \alpha \oneb^T (\Ab^{\dag} )^T \Cb_{\Smthb}^{-1} \Smthb,
\end{equation}
where
\begin{equation}
\alpha = [ \oneb^T (\Ab^{\dag})^T \Cb_{\Smthb}^{-1} \Ab^\dag \oneb]^{-1},
\end{equation}
and
\begin{equation}
\Ab^{\dag} = ( \Ab^T \Ab)^{-1} \Ab^T,
\end{equation}
and
\begin{equation}
(\Ab^{\dag} )^T = \Ab ( \Ab^T \Ab)^{-1}.
\end{equation}
Now, since it is trivially verified that $\oneb^T = \eb_1^T \Ab^T$, one obtains that 
\begin{equation}
\oneb^T (\Ab^{\dag})^T = \eb_1^T,
\end{equation}
and then, from Eq.~(\ref{eq:basic1}), $\Scmbhb = \Scmbb$. In other words, if matrix $\Cb_{\Sb}$ is
known, also in this case the ILC method provides an exact separation. When $\widehat \Cb_{\Sb}$ has to be used, 
results similar to those presented in Sect.~\ref{sec:equal} are obtained.

\subsection{Comparison with the FastIca implementation of the Independent Component Analysis (ICA)}

One conclusion that can be drawn from the arguments presented in the previous sections is that ILC suffer several drawbacks. 
However, this techniques is not the only one available for the blind separation of CMB from the Galactic foregrounds. One of the most celebrated 
competitor is the FastIca implementation of the {\it independent component analysis} (ICA). This method works explicitly under model~(\ref{eq:model}) 
with $N_o = N_c + 1$, i.e. matrix $\Ab$ is square, and noiseless data. The most interesting characteristic of ICA 
is that, unlike ILC, this technique is able to provide an estimate not only of $\Scmbb$ but also of all the Galactic components $\{ \Scbj \}$. 
For this reason, a comparison of the two methods is of interest.

The basic idea behind ICA is rather simple (and obvious): to obtain the separation of the components it is sufficient to have an estimate of matrix $\Ab$.
In fact, $\Smthb = \Ab^{-1} \Sb$. Now, if the CMB component and the Galactic ones are mutually uncorrelated (i.e. if ${\rm E}[\Smthb \Smthb^T] = \Ib$), 
then
\begin{equation} \label{eq:covA}
\Sb \Sb^T / N_p = \Ab \Ab^T.
\end{equation}
This system of equations define $\Ab$ at least of a orthogonal matrix. In fact, if $\Ab = \Zb \Vb$, with $\Vb$
orthogonal, then $\Sb \Sb^T = \Ab \Ab^T = \Zb \Vb \Vb^T \Zb^T = \Zb \Zb^T $. The problem is that, given the symmetry of $\Sb \Sb^T$,
system~(\ref{eq:covA}) contains only $N_o (N_o+1)/2$ independent equations, but the estimates of $N_o^2$
quantities should be necessary. In ICA, the $N_o (N_o-1)/2$ missing equations are obtained by imposing that the components $\Smthb$ are not only 
mutually uncorrelated but also mutually independent. In other words, the separation problem is converted into the form
\begin{equation} \label{eq:ica1}
\widehat \Smthb = \underset{\Smthb}{\arg\min} \Fb(\Smthb),
\end{equation}
\begin{equation} \label{eq:ica2}
\textrm{subject to} ~~~ \widehat \Smthb=\Ab^{-1} \Sb ~~~ \textrm{and} ~~~ \Sb \Sb^T / N_p = \Ab \Ab^T,
\end{equation}
with $\Fb(\Smthb)$ a function that measures the degree of independence between the components $\Smthb$. The definition of a 
reliable measure $\Fb(.)$ is not a trivial task. In literature, various choices are available \citep[see ][ and reference therein]{hyv01}. In practical 
algorithms, the optimization problem is not implemented explicitly in the form~(\ref{eq:ica1})-(\ref{eq:ica2}). Typically, a first estimate
$\widehat \Smthb_*$ is obtained through a {\it principal component analysis} (PCA) step followed by a {\it sphering} operation 
(i.e. normalization to unit variance). In this way,
a set of uncorrelated and normalized components are obtained, i.e. $\widehat \Smthb_* \widehat \Smthb_*^T = \Ib$. Later, this estimate
is iteratively refined to maximize $\Fb(\Smthb)$ and get the final $\widehat \Smthb$.
Again, in literature, various techniques are available. FastICA exploits a
{\it fixed-point} optimization approach \citep{hyv01,mai02,bac04}. An alternative technique is {\it JADE} that makes use of the 
{\it joint diagonalizing algorithm} \citep{car99}. 

As stated above, the main advantage of ICA is its capability to separate all the components of a given mixture, a property not shared
by ILC. Its main limit is the requirement of mutually independence of the components. This is a much stronger requirement than the 
uncorrelatedness of the the CMB templates from the Galactic ones that is required by ILC. In particular, the various
Galactic components are expected to be somewhat correlated. Another limit of some popular implementations of ICA is that the number $N_o$ 
has to be equal to $N_c+1$. If $N_o < N_c + 1$ then no solution is possible \citep[however, see][]{hyv01}. If $N_o > N_c + 1$, then the number 
$N_c$ has to be known in advance and a {\it dimension reduction} operated. In the case of noise-free observations, 
$N_c$ can be obtained from the number of non-zero eigenvalues of matrix $\widehat \Cb_{\Sb}$ that are calculated during the PCA step. Using PCA it 
is also possible to operate
the above mentioned 
{\it dimension reduction}. If noise is present, however, the identification of the non-zero eigenvalues and hence the determination of
$N_c$ may become 
difficult. This kind of problem is more serious than for ILC which does not require the exact value of $N_c$ but only that
$N_o \geq N_c + 1$.

Since problem~(\ref{eq:ica1})-(\ref{eq:ica2}) is a nonlinear one, a more detailed statistical characterization of the ICA properties represents 
a quite difficult task. In any case, from the considerations above, one has not to expect that ICA could outperform ILC. On the other hand, 
given its important limits, ILC cannot be expected to offer much better performances. 

\subsection{Some Numerical Experiments}

To support the results numerical experiments are presented here. In these simulations we have
deliberately chosen non-astronomical ``deterministic'' subjects. The reason is that in this way
a direct visualization of the separation provided by ILC and ICA is possible as well as a simpler modeling of particular experimental conditions 
(e.g., the spurious correlation between different images due to their finite sizes). The fact that in the previous sections
the CMB as well as the Galactic components are assumed to be the realization of stationary random processes, does
not represent a limit since the arguments have been developed assuming fixed realizations of those processes.
This permits to interpret the separation as an operation in a deterministic framework. The only difference is represented by the fact that, for genuine 
deterministic signals, the various statistical quantities and operators lose their statistical meaning. For example, the cross-covariance matrix 
$\widehat \Cb_{\Sb}$ coincides with $\Cb_{\Sb}$ and becomes a measure of the coherence between signals. Something similar holds for the
other operators.
 
\begin{figure*}
        \resizebox{\hsize}{!}{\includegraphics{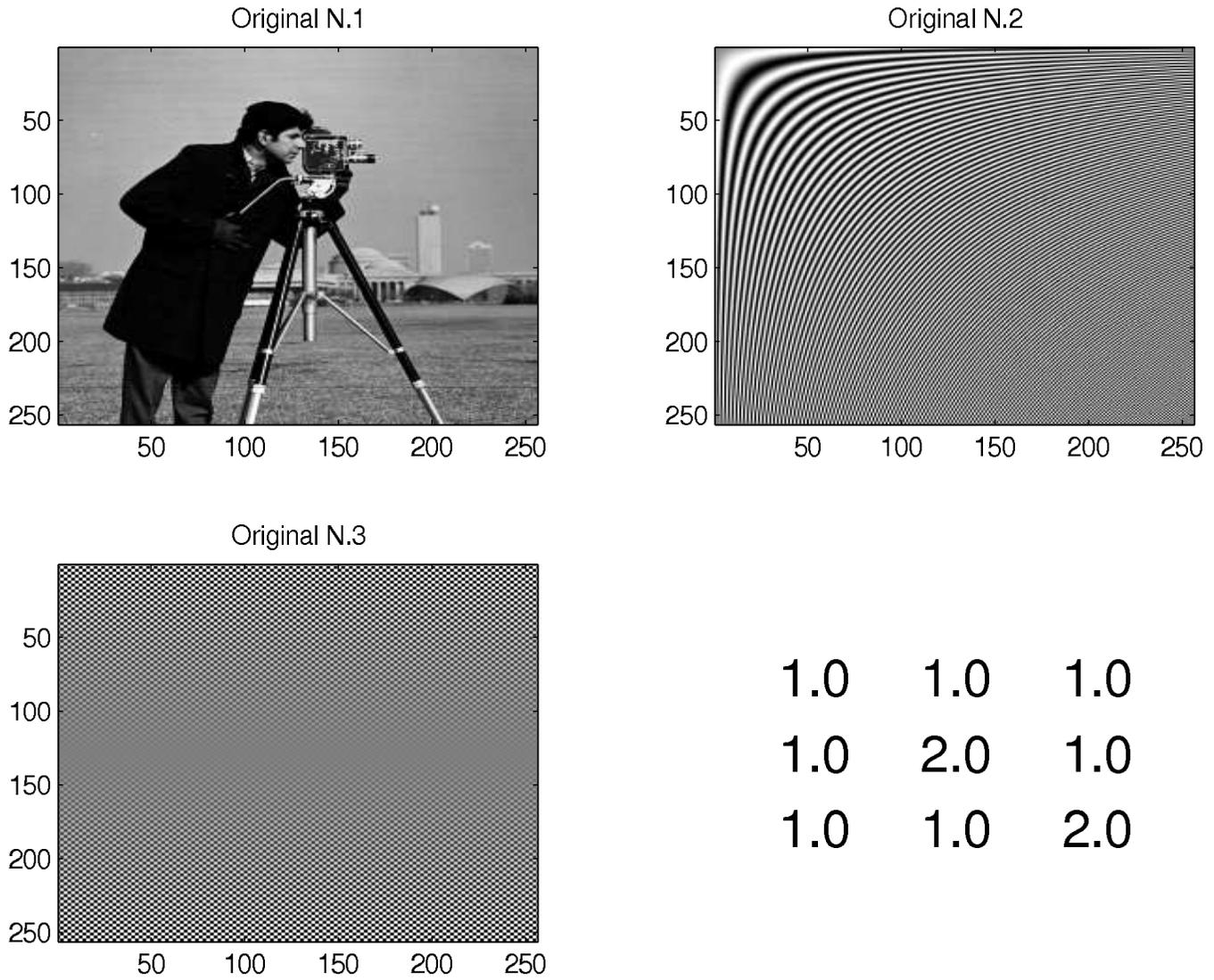}}
        \caption{Original images $\Smthb$ -- see Eq.~(\ref{eq:model}) --
        used in the first experiment described in the text (i.e., separation of almost uncorrelated components) 
        to test both ILC and ICA (FastICA algorithm) techniques. The top-left panel shows the image to recover. All the images, before 
        the mean subtraction, have values in the interval $[0, 1]$. The bottom-right panel provides the mixing matrix $\Ab$ -- see 
        Eq.~(\ref{eq:Amatrix}).}
        \label{fig:original2}
\end{figure*}

\begin{figure*}
        \resizebox{\hsize}{!}{\includegraphics{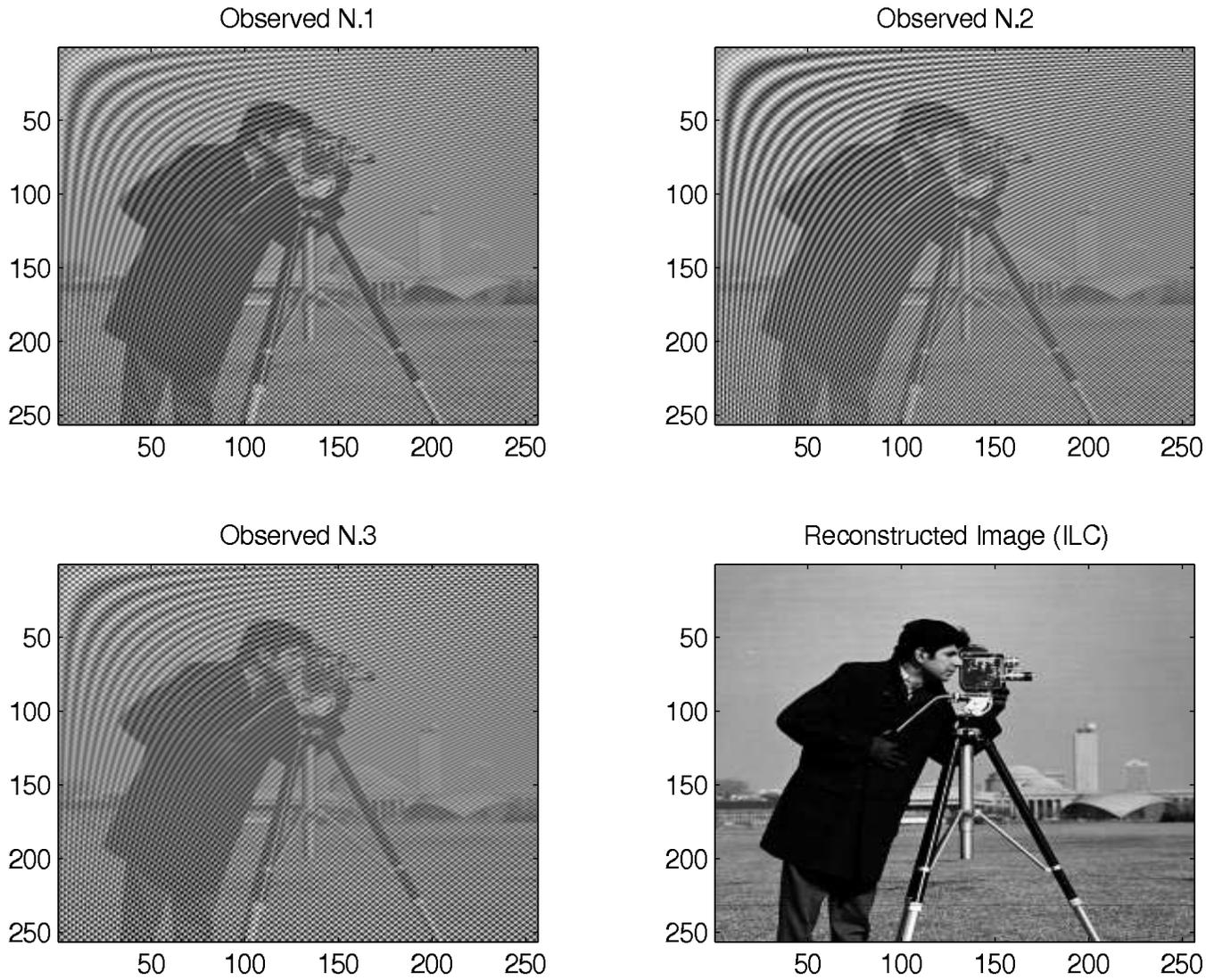}}
        \caption{The three mixtures (observed images) $\Sb = \Ab \Smthb$ used in the first experiment to test the ILC and ICA 
        (FastICA algorithm). Here, the experiment mimics three independent observing channels and the separation
        in the case of three almost uncorrelated components. The bottom-right panel displays the resulting ``restored'' 
        image obtained with ILC. As seen the ILC method retrieves the input image quite well.}
        \label{fig:observed2}
\end{figure*}

\begin{figure*}
        \resizebox{\hsize}{!}{\includegraphics{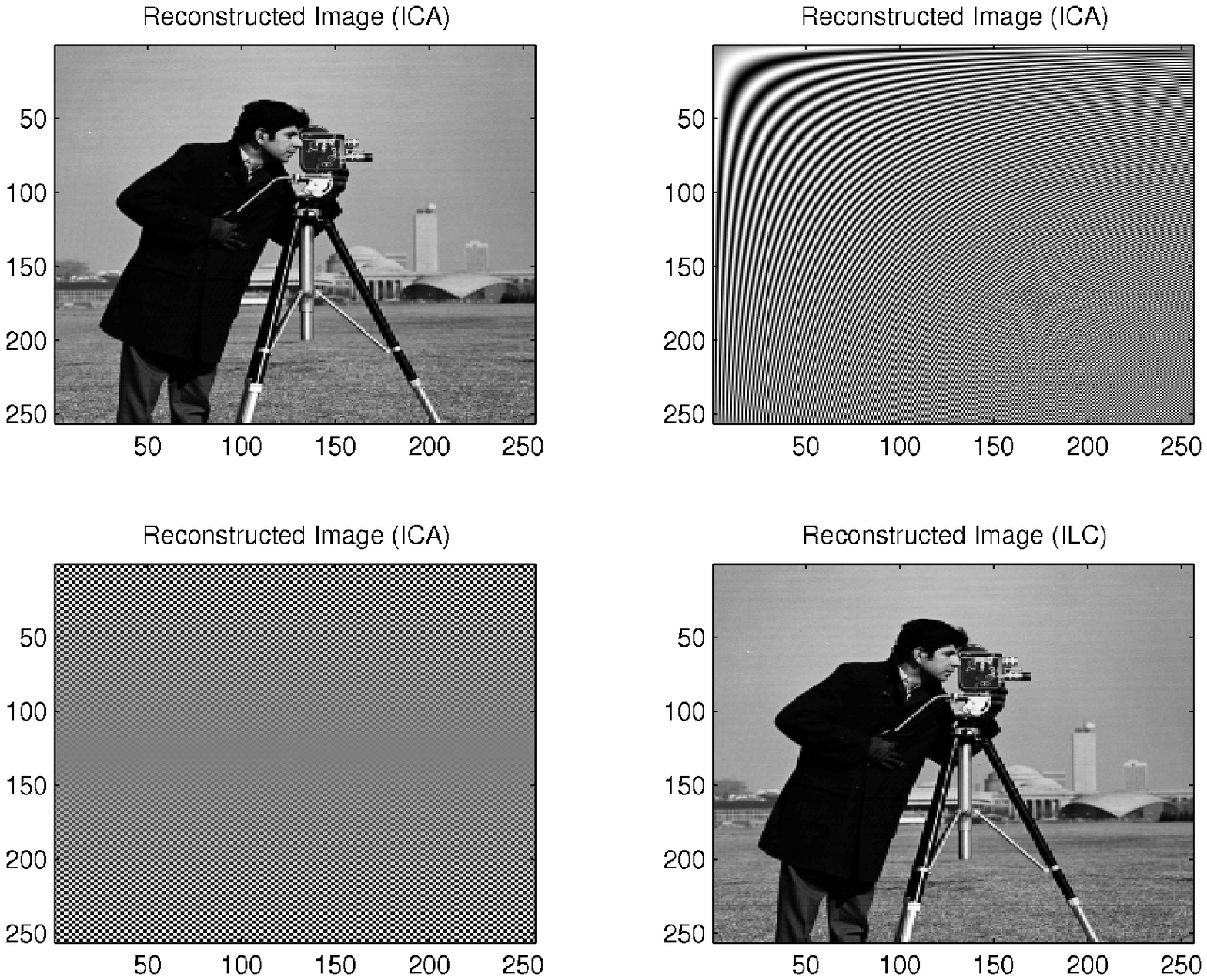}}
        \caption{Comparison between the results obtained with ICA (FastICA algorithm) and ILC concerning the mixtures in Fig.~\ref{fig:observed2}.
        The first three panels show the three components reconstructed by the ICA algorithm. The bottom right panel shows
        the result obtained with ILC, already shown in Fig.~\ref{fig:observed2}.}
        \label{fig:restored2}
\end{figure*}

\begin{figure*}
        \resizebox{\hsize}{!}{\includegraphics{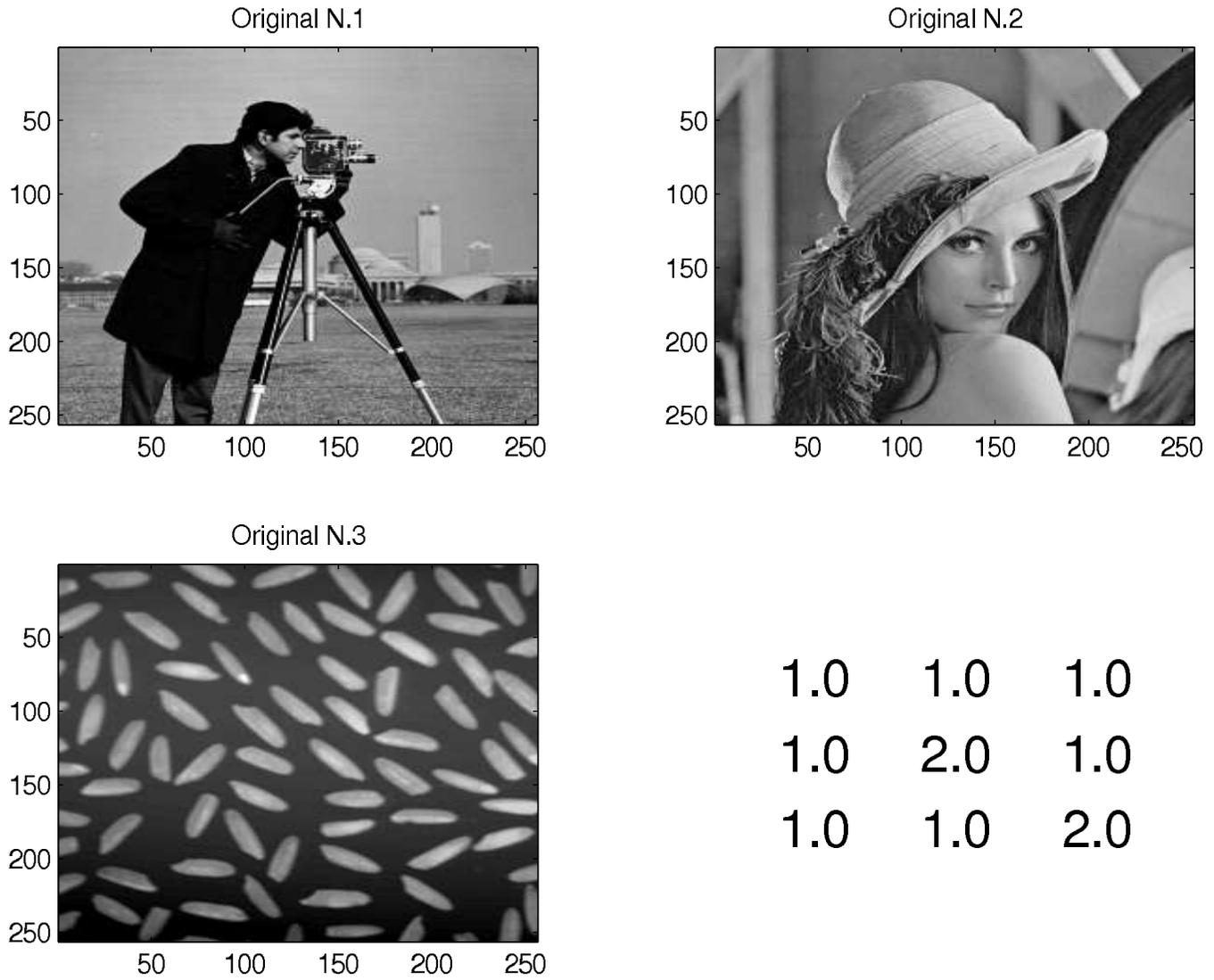}}
        \caption{Original images $\Smthb$ used in the second experiment described in the text (i.e., separation of partially correlated components)
        to test the ILC and ICA (FastICA algorithm) techniques. The top-left panel shows the image to recover.
	  Here a spurious correlations among the images is introduced (as often the case in foreground components of the CMB).
        These images before the mean subtraction, have values in the interval $[0, 1]$.
        The bottom-right panel provides the mixing matrix $\Ab$.}
        \label{fig:original1}
\end{figure*}

\begin{figure*}
        \resizebox{\hsize}{!}{\includegraphics{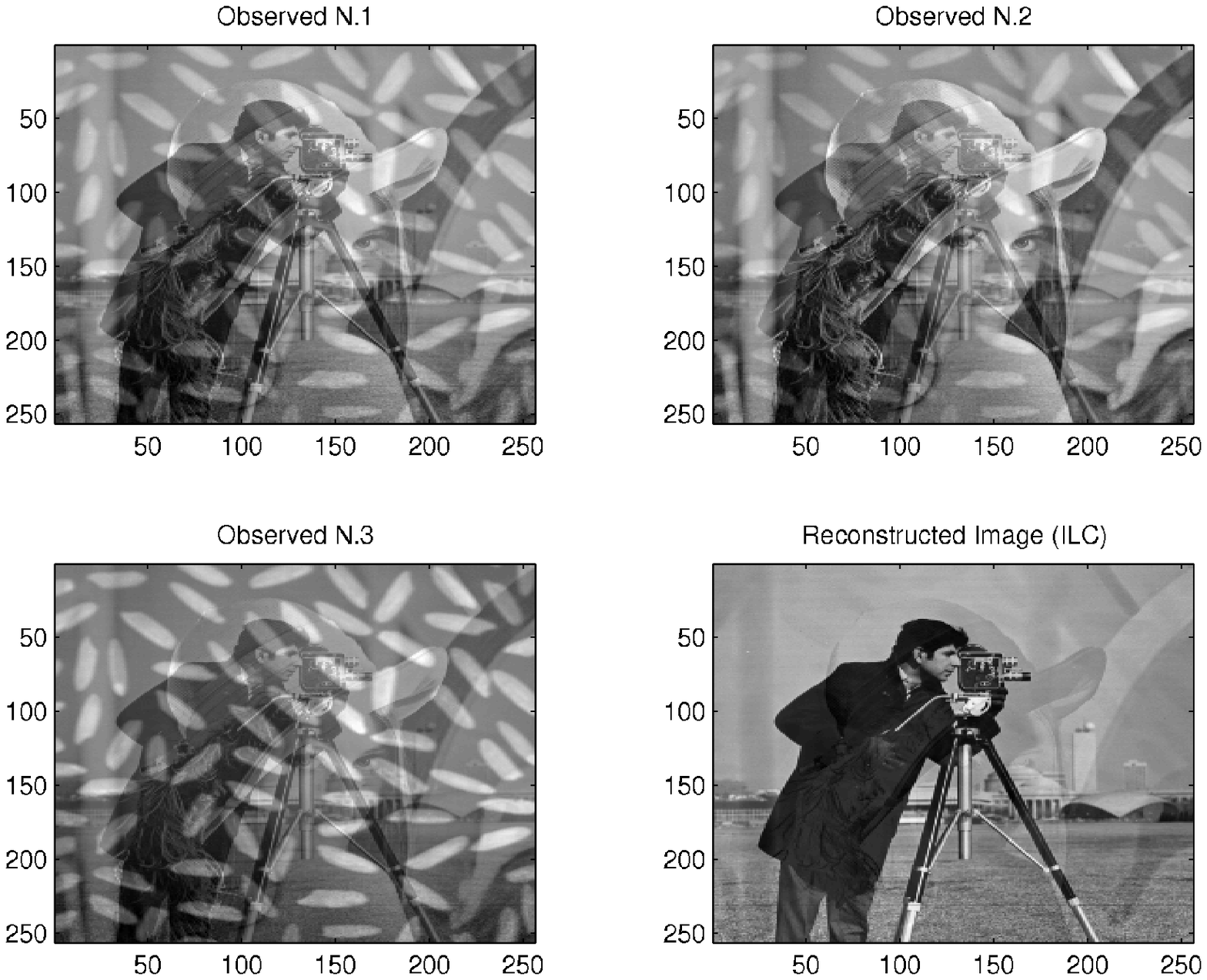}}
        \caption{The three mixtures (observed images) $\Sb = \Ab \Smthb$ used in the second experiment to test the ILC and ICA 
        (FastICA algorithm).  Here, the experiment mimics three independent observing channels and the separation
        in the case of three almost uncorrelated components. The bottom-right panel displays the resulting ``restored'' 
        image obtained with ILC. As seen the method is not able to fully retrieve the input image.}
        \label{fig:observed1}
\end{figure*}

\begin{figure*}
        \resizebox{\hsize}{!}{\includegraphics{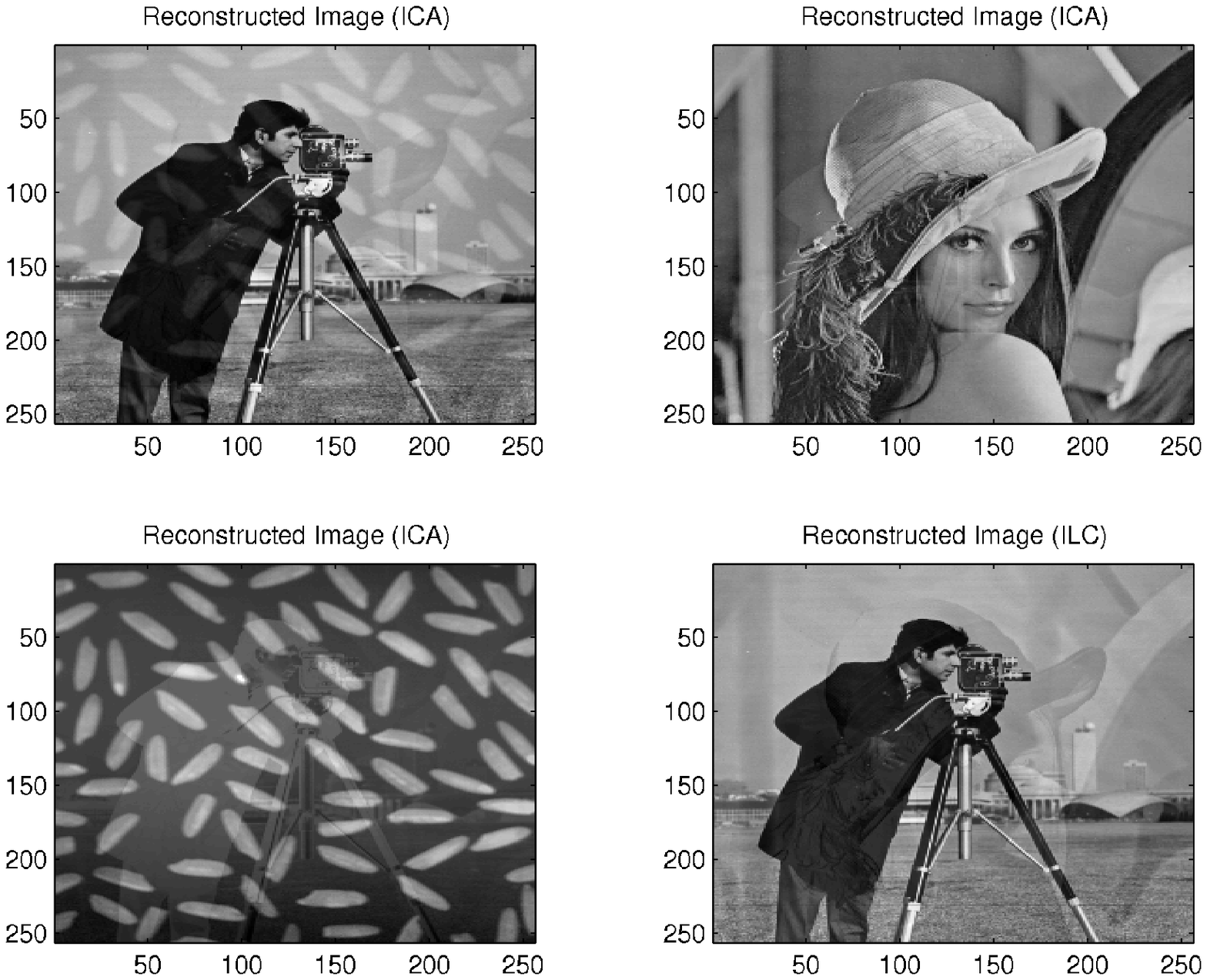}}
        \caption{Comparison between the results obtained with ICA (FastIca algorithm) and ILC concerning the mixtures in Fig.~\ref{fig:observed1}.
        The first three panels show the three components reconstructed by the ICA algorithm. The bottom right panel shows
        the result obtained with ILC, already shown in Fig. ~\ref{fig:observed1}. This figure has to be compared with Fig.~\ref{fig:restored2}.}
        \label{fig:restored1}
\end{figure*}

In the first experiment, whose results are shown in Figs.~\ref{fig:original2}-\ref{fig:restored2}, the performance of both ILC and ICA is tested 
under favorable conditions. 
In particular, three mixtures (Fig.~\ref{fig:observed2}) have been simulated through model~(\ref{eq:model}) using an equal number of 
component images (Fig.~\ref{fig:original2}) 
that are almost uncorrelated with each other. The mixing matrix $\Ab$~(\ref{eq:Amatrix}) is given in Fig.~\ref{fig:original2}. The quality of 
the separation in Fig.~\ref{fig:restored2} is excellent. This result can be ascribed to the small values of 
the off-diagonal entries of the cross-covariance matrix as supported by Figs.~\ref{fig:original1}-\ref{fig:restored1} that present the results
of an experiment similar to the previous one. Here, however, a spurious dependence between the components of the mixtures is
mimicked using three images having almost-constant luminosity areas in correspondence to the 
same coordinates (Fig.~\ref{fig:original1}) . Such a spurious dependence is highlighted by the (normalized) cross-covariance matrix 
$\widehat \Cb_{\Smthb}$
\begin{equation}
\widehat \Cb_{\Smthb} = 
\left( \begin{array}{ccc}
1.00 &  0.20 & 0.06 \\
0.20 &  1.00 & 0.06 \\
0.06 &  0.06 & 1.00
\end{array} \right).
\end{equation}
The off-diagonal entries are much larger than zero and the separation provided Fig.~\ref{fig:restored1} by both the methods is undoubtedly worse.

\begin{figure*}
        \resizebox{\hsize}{!}{\includegraphics{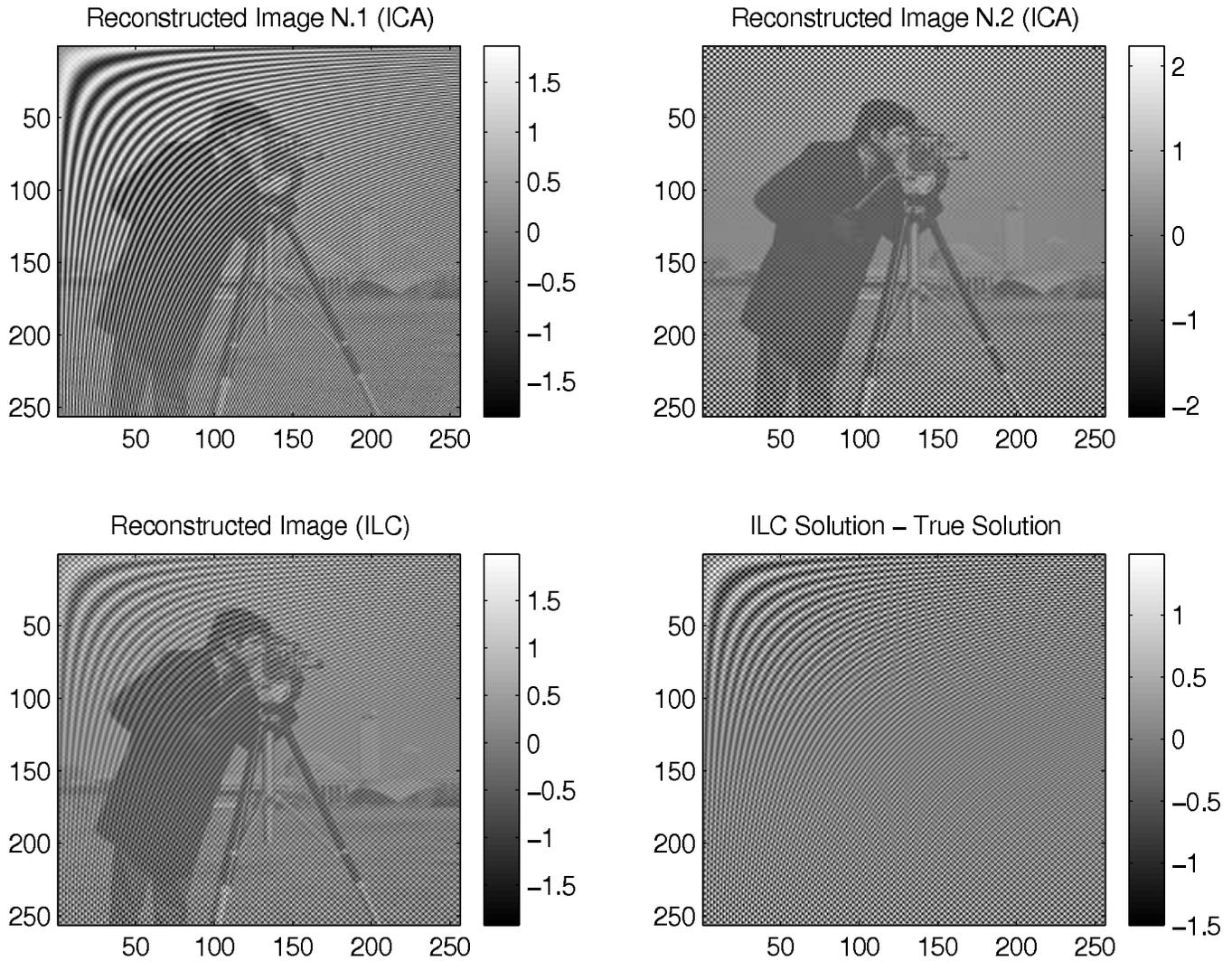}}
        \caption{Resulting images as in the experiment shown in Figs.~\ref{fig:original2}-\ref{fig:restored2}. In this case the number of components
         affecting the signal is larger than the number of mixtures (observed images). The case shown in this figure corresponds
         to three original input images and only two mixtures available. 
         The bottom-right panel shows the residuals, $\Delta \Scmbb = \Scmbhb - \Scmbb$, in the ILC method. These residuals are of the same order of
         magnitudes as the input images and therefore the method is not able to retrieve the original components.
         This figure has to be compared with Fig.~ \ref{fig:restored2}.}
        \label{fig:biased}
\end{figure*}

The aim of the last experiment is to test the result of Sec.~\ref{sec:minus} that ILC is not able to provide 
reliable results when the number of observed mixtures is smaller than the number of the corresponding components. In this respect,
Fig.~\ref{fig:biased} offers an illuminating example.
This experiment is again similar to the one corresponding to Figs.~\ref{fig:original2}-\ref{fig:restored2}, with the difference that only two mixtures 
are available with the mixing matrix
\begin{equation}
\Ab = 
\left( \begin{array}{ccc}
1.0 & 2.0 & 1.0 \\
1.0 & 1.0 & 2.0 \\
\end{array} \right).
\end{equation}
The different quality of the separation does not deserve comment. Indeed, from Eq.~(\ref{eq:bias}) it is
\begin{equation}
\Delta \Scmbb = (0, 1.19, 1.87) \Smthb,
\end{equation}
i.e. a quantity whose magnitude is comparable to the image of interest (see also the bottom-right panel in Fig.~\ref{fig:biased}). 
Not unexpectedly, ICA as well seems to be unable to provide an acceptable separation.

\section{STATISTICAL CHARACTERISTICS OF ILC: NOISY OBSERVATIONS}
\label{sec:ILCnoise}

\begin{figure*}
        \resizebox{\hsize}{!}{\includegraphics{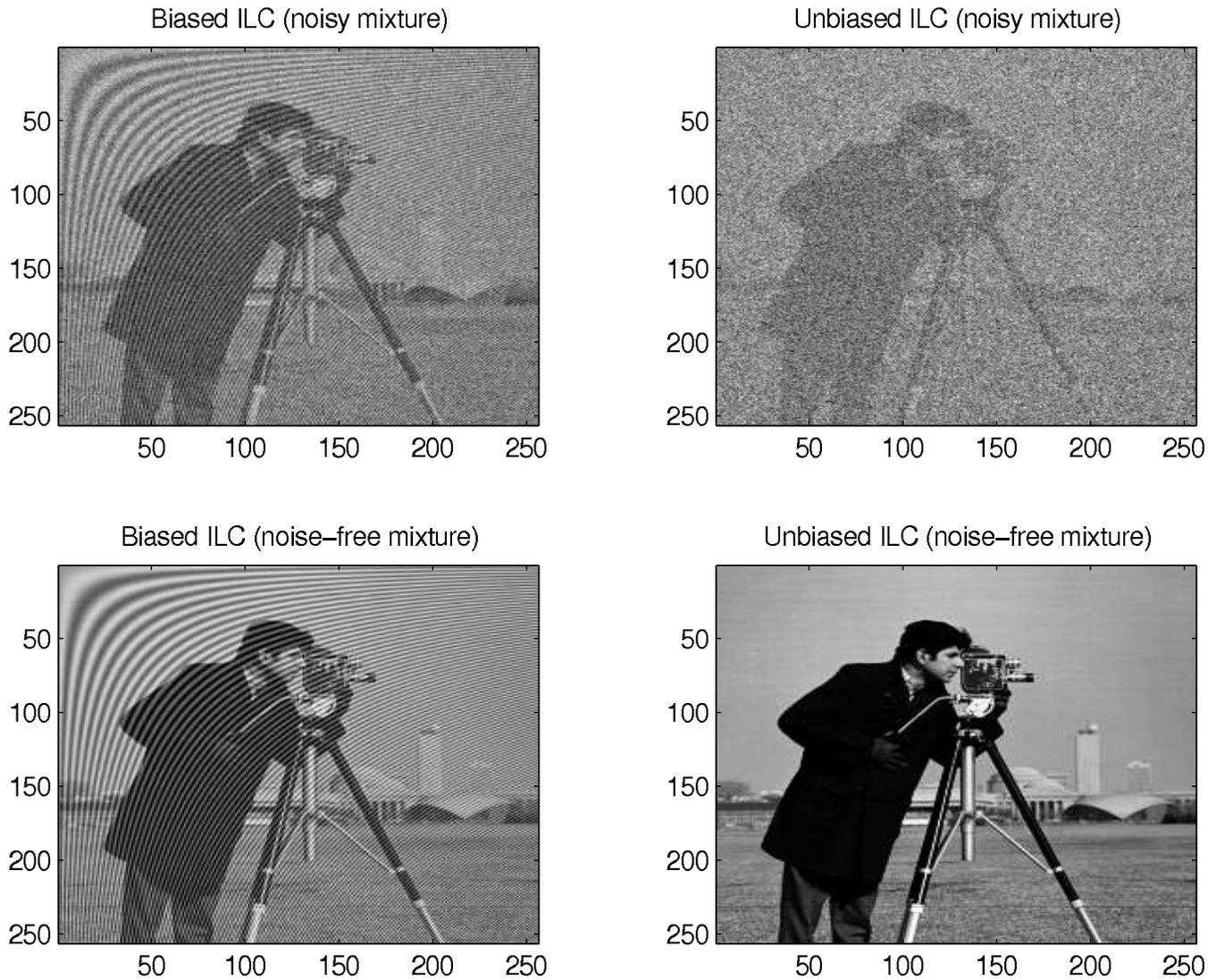}}
        \caption{The figure shows the results of noise effects for the biased, respectively, unbiased ILC estimator as described in the text 
        (see Sec.~\ref{sec:ILCnoise}): {\bf Top panels} -- Comparison between the results obtained with the two estimators
        when to the mixtures a Gaussian, zero-mean, white-noise process with ${\rm SNR}=20$ is added. The presence of the bias for the
        biased estimator and the amplification of the noise for the unbiased one are evident.
        {\bf Lower panels} -- Results obtained when the same estimators are applied to the
        noise-free version of the mixtures. This is to better characterize the bias introduced by the biased estimator and the unbiasedness of the
        other one.}
        \label{fig:noise1}
\end{figure*}

In situations of maps contaminated by measurement noise $\Nmthb$, assumed zero-mean, additive, and stationary, model~(\ref{eq:model}) becomes
\begin{equation} \label{eq:modeln}
\Sb = \Ab \Smthb + \Nmthb.
\end{equation}
In turn, under the condition of $\Nmthb$ uncorrelated with $\Sb$, $\Cb_{\Sb}$ becomes
\begin{equation} \label{eq:covar}
\Cb_{\Sb} = \Ab \Cb_{\Smthb} \Ab^T + \Omegab_{\Nmthb},
\end{equation}
with
\begin{equation}
\Omegab_{\Nmthb} = {\rm E}[ \Nmthb \Nmthb^T].
\end{equation}
If noise is small and $N_o = N_c + 1$, when Eq.~(\ref{eq:covar}) is inserted in Eq.~(\ref{eq:basic}) and the resulting expression expanded
in Taylor series up to the linear term, it is possible to see that
\begin{equation} \label{eq:result}
\Scmbhb = \alpha^* [\oneb^T - \gammab] \Ab^{-T} \Cb_{\Smthb}^{-1} \Smthb,
\end{equation} 
with $\gammab = \sigma_{cc}^{-1} \eb_1^T \Ab^{-1} \Omegab_{\Nmthb}$ and $\alpha^*$ a constant coefficient 
that depends on $\Omegab_{\Nmthb}$ and converges to $\alpha$ for decreasing level of the noise. 
A similar result holds when $N_o > N_c + 1$ with $(\Ab^{\dag})^T$ substituting $\Ab^{-T}$.
From Eq.~(\ref{eq:result}) it is clear that ${\rm E}[\Scmbhb | \Ab \Sbb  ] \neq \Scmbb$, i.e. the ILC estimator is biased.
Since $\gammab$ depends on the specific characteristics of $\Ab$ and $\Omegab_{\Nmthb}$, this does not allow a general
treatment of the question. However, it is not difficult to realize that the bias affecting $\Scmbhb$ can be severe 
(e.g., see the top-left panel of Fig.~\ref{fig:noise1}).

It is possible to obtain an unbiased
ILC estimator computing the weights $\wb$ by means of Eq.~(\ref{eq:basic}) with the cross-covariance matrix $\Cb_{\Sb}$ substituted by
\begin{equation} 
\Cb_{\Sb}^* = \Cb_{\Sb} - \Omegab_{\Nmthb}.
\end{equation}
This operation, however, has a cost: the weights $\wb$ do not minimize any longer
the variance of $\{ \Scmbh \}$. As a consequence, the influence of noise can be even dramatically amplified. This fact is clearly visible in the 
top panels of Figs.~\ref{fig:noise1} that compare the results obtained by the ILC estimator based on $\widehat \Cb_{\Sb}$ and 
$\widehat \Cb_{\Sb}^* = \widehat \Cb_{\Sb} - \Omegab_{\Nmthb}$, respectively.
Two mixtures have been simulated using the two figures in the top panels 
of Fig.~\ref{fig:original2} and the mixing matrix
\begin{equation}
\Ab = 
\left( \begin{array}{cc}
1.00 & 1.00 \\
1.00 & 0.95
\end{array} \right).
\end{equation}
A Gaussian, white-noise process has been added to these mixtures with a signal-to-noise-ratio (${\rm SNR}$) \footnote{Here, ${\rm SNR}$ is defined as the
ratio of the standard deviation of the signal with respect to the standard deviation of the corresponding noise.} equal to 20. 
The reason of such unpleasant result has to be searched in the condition number, $\kappa(\Cb_{\Sb}^*)$, of $\Cb_{\Sb}^*$ that tends to be larger 
than $\kappa(\Cb_{\Sb})$. Because of this, the entries of $(\Cb_{\Sb}^*)^{-1}$ can take values in a range much wider than those of $\Cb_{\Sb}^{-1}$ and
the same holds for the corresponding weights $\wb$. In fact, according to a theorem due to Weyl \citep[e.g., see ][Theorem III.2.2]{bha97}, 
it is
\begin{align}
\lambda_1(\Cb_{\Sb}) - \lambda_1(\Omegab_{\Nmthb}) & \le \lambda_1(\Cb_{\Sb}^*) \le \lambda_1(\Cb_{\Sb}) - \lambda_{N_o}(\Omegab_{\Nmthb}), \\
\lambda_{N_o}(\Cb_{\Sb}) - \lambda_1(\Omegab_{\Nmthb}) & \le \lambda_{N_o}(\Cb_{\Sb}^*) \le \lambda_{N_o}(\Cb_{\Sb}) - \lambda_{N_o}(\Omegab_{\Nmthb}),
\end{align}
where $\lambda_i(\Hb)$ denotes the i-th eigenvalues of a $N \times N$ symmetric, positive definite matrix $\Hb$ with $ \lambda_1 \ge \lambda_2 \ldots 
\ge \lambda_{N} > 0 $.
These inequalities indicate an high probability that $\lambda_{N_o}(\Cb_{\Sb}^*) / \lambda_{N_o}(\Cb_{\Sb}) <
\lambda_1(\Cb_{\Sb}^*) / \lambda_1(\Cb_{\Sb})$. In this way,
\begin{equation}
\kappa(\Cb_{\Sb}^*) = \frac{\lambda_1(\Cb_{\Sb}^*)}{\lambda_{N_o}(\Cb_{\Sb}^*)} > 
\frac{\lambda_1(\Cb_{\Sb})}{\lambda_{N_o}(\Cb_{\Sb})} = \kappa(\Cb_{\Sb}).
\end{equation}
For example, this is strictly true if $\Omegab_{\Nmthb}$ is a diagonal matrix with non-zero entries equal to a 
constant value $\varrho$, since 
\begin{equation}
\kappa(\Cb_{\Sb}^*) = \frac{\lambda_1(\Cb_{\Sb}) - \varrho}{\lambda_{N_o}(\Cb_{\Sb}) - \varrho}.
\end{equation}
From these consideration, in the case of low ${\rm SNR}$ and large  $\kappa(\Cb_{\Sb}^*)$, one has to expect important amplification of 
the noise for the unbiased ILC estimator. 

\begin{figure}
        \resizebox{\hsize}{!}{\includegraphics{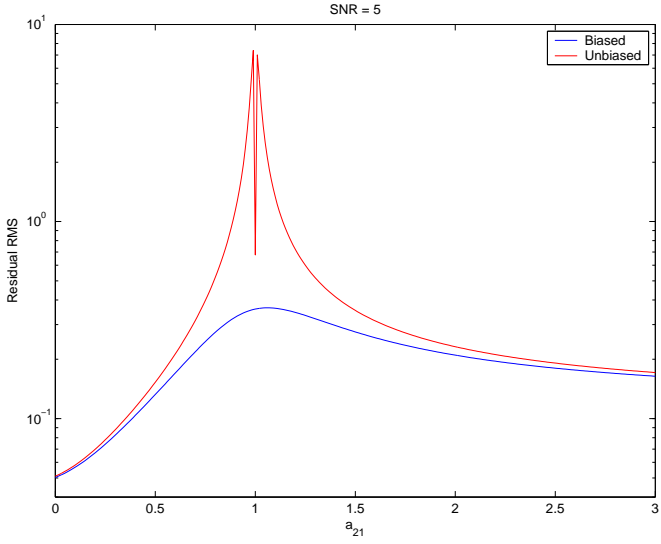}}
        \caption{Experiment showing the effect of noise: RMS of the residuals for the biased, respectively, unbiased ILC estimators vs. 
        $a_{21}$ when ${\rm SNR} = 5$. Here, $a_{21}=1$ corresponds to a singular mixing matrix $\Ab$. The dip in correspondence to this value
        is a numerical effect do to the use of the sample $\widehat \Cb_{\Sb}^*$.}
        \label{fig:rms_5}
\end{figure}

\begin{figure}
        \resizebox{\hsize}{!}{\includegraphics{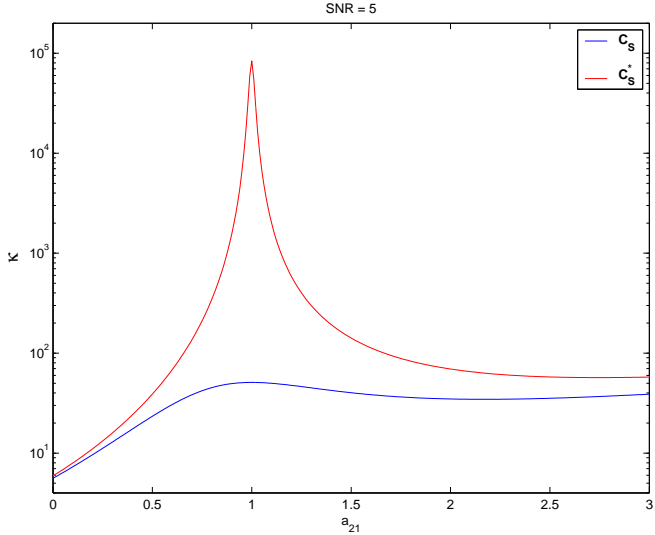}}
        \caption{Experiment concerning the influence of the noise: condition number $\kappa$ of $\widehat \Cb_{\Sb}$, respectively, 
        $\widehat \Cb_{\Sb}^*$ vs. $a_{21}$ when ${\rm SNR} = 5$. Here, $a_{21}=1$ corresponds to a singular mixing matrix $\Ab$.}
        \label{fig:cond_5}
\end{figure}

\begin{figure}
        \resizebox{\hsize}{!}{\includegraphics{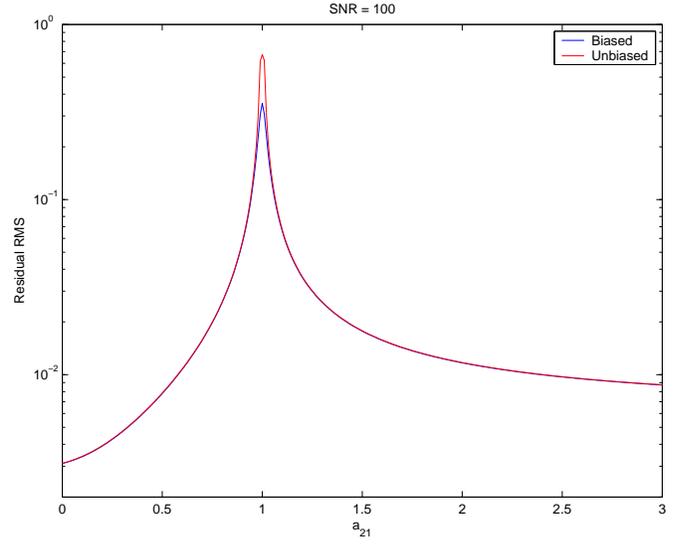}}
        \caption{Experiment showing the effect of noise: RMS of the residuals for the biased, respectively, unbiased ILC estimators vs. 
        $a_{21}$ when ${\rm SNR} = 100$. Here, $a_{21}=1$ 
        corresponds to a singular mixing matrix $\Ab$.}
        \label{fig:rms_100}
\end{figure}

\begin{figure}
        \resizebox{\hsize}{!}{\includegraphics{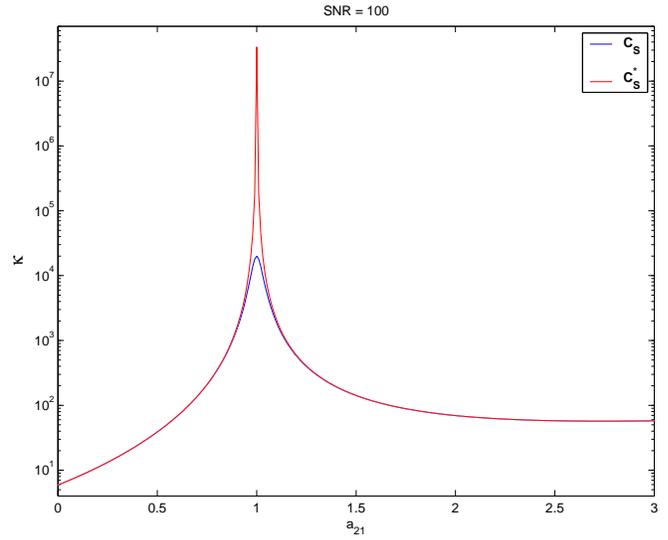}}
        \caption{Experiment showing the effect of noise: condition number $\kappa$ of $\widehat \Cb_{\Sb}$, respectively, 
        $\widehat \Cb_{\Sb}^*$ vs. $a_{21}$ when ${\rm SNR} = 100$. Here, $a_{21}=1$ corresponds to a singular mixing matrix $\Ab$.}
        \label{fig:cond_100}
\end{figure}

The correctness of this argument is supported 
by Figs.~\ref{fig:rms_5}-\ref{fig:cond_100} that regard an experiment similar to that corresponding to Fig.~\ref{fig:noise1}. 
These figures show the root-mean-square (rms) of the residuals corresponding to the biased, respectively, unbiased
ILC estimators for ${\rm SNR}= 5, 100$ and various values of $\kappa$. Different values of this last parameter has been obtained assuming
a mixing matrix with the form
\begin{equation}
\Ab = 
\left( \begin{array}{cc}
1.00 & 1.00 \\
1.00 & a_{21}
\end{array} \right),
\end{equation}
and making $a_{21}$ to assume a set of evenly spaced values in the range $[0, 3]$. When $a_{21}=1$, matrix $\Ab$ becomes singular. An interesting 
indication that come out from these figures is that,
in the case of high ${\rm SNR}$ the unbiased ILC estimator provides remarkably worse result only when matrix $\Cb_{\Sb}^*$ is very ill-conditioned.
This may happen if the different images are very similar (observations performed at very similar frequencies). This
suggests that, if $\kappa(\Cb_{\Sb}^*)$ is not too large, the unbiased ILC estimator could be used with the benefit that the 
filtering a noise is typically an easier operation than the removal of a bias. In any case, even in the case $\Cb_{\Sb}^*$ be ill-conditioned,
as it happens when $N_o > N_c +1$, its $\kappa$ can be reduced by means of the SVD technique presented in Sec.~\ref{sec:more}.

It is necessary to stress that all of these conclusions hold only in situations of isotropic noise
(i.e., a noise with identical characteristics everywhere). In the contrary case, 
bad results have to be expected since noise behaves as a sort of additional channel-dependent component.

\section{CONCLUSIONS} \label{sec:conclusions}

The arguments presented in the previous sections show that for a safe use of the ILC estimator some conditions have to be satisfied.
In particular:
\begin{enumerate}
\item The observations have to cover a sky area much wider than the spatial scale (correlation length) of the observed maps in such a way to
allow an accurate estimate of the cross-covariance matrix $\Cb_{\Sb}$;
\item Model~(\ref{eq:model})-(\ref{eq:Amatrix}) has to hold. In particular, ILC cannot be expected to produce satisfactory results 
if some of the Galactic templates depend on the observing channels. This point can be realized
if a Galactic channel-dependent template $\Sbb_j$ is thought as the sum of $N_j$ channel-independent templates. 
In this case, model~(\ref{eq:model})-(\ref{eq:Amatrix}) still holds but with an effective number of Galactic components that now is 
$N_c^* = \sum_{j=1}^{N_c} N_j$. 
Since the number of observing channel is typically rather limited, in practical applications it has to be expected 
that $N_o < N_c^* + 1$ and then the point below applies;
\item The condition $N_o \ge N_c^* + 1$ has to hold, i.e. the number of the observing channels has to be equal or larger than the number of the
physical components (CMB included) that contribute to form the observed maps. In the contrary case, the solution can suffer a severe distortion;
\item The level of the noise has to be rather low otherwise a severe bias can be introduced in the ILC estimator.  
This can be removed but at the price of a possible remarkable amplification of the noise influence. However, especially in situations of 
high ${\rm SNR}$, the use of the unbiased ILC estimator can offer some advantages.
\end{enumerate}
From these points it is not difficult to realize that the ILC estimator is not trivial to use. In particular,
the first two points conflict each other. In fact, in the case of wide maps,   
model~(\ref{eq:model})-(\ref{eq:Amatrix}) is not applicable since on large spatial scales the Galactic templates are expected to depend on the 
observing channel. In this respect, a simple tests can be of help that is based on the analysis of the eigenvalues of 
$\widehat \Cb_{\Sb}$. In fact, if $N_o \geq N_c^* + 1$, matrix $\widehat \Cb_{\Sb}$ has to be (almost) singular with a 
number of (almost) zero elements in the diagonal matrix $\widehat \Db$ of Eq.(\ref{eq:svd}) equal to $N_o - N_c^* - 1$. Unfortunately, this
test can be thwarted by the noise; because of
the statistical fluctuations, the entries of $\widehat \Db$ that should be close to zero can take larger value. 

The obvious conclusion is that, in order to obtain reliable results, the use of ILC requires a careful planning of the observations. 
The area of the sky to observe as well as the tolerable level of the noise are factors that have to be fixed in advance. To try an
``{\it a posteriori}'' correction of the distortions introduced in the ILC solution by the violation of the above conditions is a quite risky operation. 
The question is that all of these distortions critically depend on the true solution that one tries to estimate and thereof they cannot be obtained 
from the data only. The alternative represented by the numerical simulations that make use of semi-empirical templates is quite risky. In particular, 
there is the concrete possibility to force spurious features in the final results. This holds also in the case these results are used as ``{\it prior}''
in more sophisticated separation techniques.

A last comment regards the use ILC in the Fourier domain. Since this method provides the same result independently of the domain, it could
seem that there is no particular benefit to work in the Fourier one. Actually, this can be not true if the maps have different spatial resolutions
(i.e. the observing channels have different point spread functions) and/or whether a frequency-dependent separation is desired \citep{teg03}. 
Although in this way it is possible to improve some properties of the separated maps (e.g., the spatial resolution), it is necessary to stress that 
this has a cost in the amplification of the noise level. 

\begin{acknowledgements}
We thank Dr. Carlo Baccigalupi for careful reading of this manuscript.
\end{acknowledgements}

\clearpage
\clearpage
\clearpage
\clearpage
\clearpage
\clearpage
\clearpage
\clearpage

\end{document}